# Plasmonic Parametric Absorbers

SHIMA FARDAD,[1,2] ALESSANDRO SALANDRINO[1,2,*]

[1]*Department of Electrical Engineering and Computer Science, University of Kansas, 1520 West 15th Street, Lawrence, Kansas 66045, USA*
[2]*University of Kansas, ITTC, 2335 Irving Hill Road, Lawrence, Kansas 66045, USA*
*\*Corresponding author: a.salandrino@ku.edu*

**Exploiting the dynamics of Plasmonic Parametric Resonance (PPR), we introduce the theory of Plasmonic Parametric Absorbers (PPA). The key insight informing the PPA idea is that in the PPR process a pump field experiences an extinction rate that strongly depends on the intensity of the pump itself, creating two distinct regimes: one of weak absorption under low intensity illumination, and one of strong absorption when the threshold of parametric resonance is met or exceeded. Due to this reverse saturable absorption behavior, PPAs are promising candidates for optical limiting applications.**

Absorption, which is arguably the Achilles' heel of most metamaterials applications, in the case of absorbers is a desired effect that can be accurately tailored exploiting the properties of nanostructured media. Since the seminal work of Landy et Al.[1] a great deal of attention has been devoted to the design of metamaterial absorbers [2] operating in a variety of spectral regions, from the terahertz range [3] up to the optical domain [4]. The degrees of freedom afforded by metamaterial structures have led to the realization of absorbers with unprecedented characteristics in terms of efficiency, spectral and/or angular bandwidth [5-8] or selectivity [9, 10]. Nonlinearity can greatly enrich the optical absorption phenomenology, and a host of applications can be envisioned exploiting the unique properties of nonlinear metamaterial absorbers, ranging from all-optical modulation schemes to optical-limiting devices.

Here we present the theory of a new class of nonlinear absorbers termed Plasmonic Parametric Absorbers (PPA) relying on the recently introduced concept of Plasmonic Parametric Resonance (PPR) [11]. In contrast with conventional localized plasmonic resonances, whereby modes are excited directly by an external field of frequency and spatial profile matching those of a given mode of the plasmonic particle, PPR is a form of amplification in which a pump field transfers energy to a mode in an indirect fashion. In PPR in fact the modes of a plasmonic structure are amplified by means of a temporal permittivity modulation of the background medium interacting with an appropriate pump field. Such permittivity variation translates into a modulation of the modal resonant frequency, and under specific conditions amplification can occur. As shown in [11], among the unique characteristics of PPR is the possibility of accessing modes of arbitrarily high order with a simple spatially uniform pump, provided that such pump exceeds a certain intensity threshold. As we will show in the following, it is such threshold behavior that can lead to a type of nonlinear metamaterial absorber with rather unique properties.

The key insight informing the PPA idea is that in the PPR process the pump field experiences an extinction rate that strongly depends on the intensity of the pump itself, creating two distinct regimes: one of weak absorption under low intensity illumination, and one of strong absorption when the threshold of parametric resonance is met or exceeded. The characterization of the pump field extinction is the main focus of our theoretical analysis. In addition, the theory of PPR is extended here to predict and quantify the effects of saturation dynamics in the high-intensity/high-absorption regime. The system under consideration is a composite medium of identical subwavelength plasmonic particles dispersed in a dielectric host material. An electric field $\mathbf{E}_P(t)$, which will be referred to as "pump", is incident on the system. It is assumed that the electromagnetic response of the medium constituting the particles (henceforth referred to as "medium 1") is dominated by free-carriers effects. Under such assumption the dispersive component of the polarization density $\mathbf{P}_1(\mathbf{r},t)$ in medium 1 evolves in time according to the following differential equation:

$$\frac{\partial^2 \mathbf{P}_1(\mathbf{r},t)}{\partial t^2} + \gamma \frac{\partial \mathbf{P}_1(\mathbf{r},t)}{\partial t} = \varepsilon_0 \omega_{pl}^2 \mathbf{E}_1(\mathbf{r},t) , \quad (1)$$

where $\mathbf{E}_1(\mathbf{r},t)$ is the electric field within medium 1. It is worth noting that equation (1) is equivalent to assigning to medium 1 a Drude-like permittivity with plasma frequency $\omega_{pl}$ and collision frequency $\gamma$. For completeness a nondispersive permittivity term $\varepsilon_\infty$ will be also included to account for higher frequency spectral features of the dielectric response of medium 1. The background medium, henceforth "medium 2", is assumed non-dispersive with relative permittivity $\varepsilon_2$ and endowed with second order nonlinearity $\chi^{(2)}$. Under such hypotheses the polarization density $\mathbf{P}_2(\mathbf{r},t)$ in medium 2 can be expressed in terms of the total local field $\mathbf{E}_2(\mathbf{r},t)$ as follows:

$$\mathbf{P}_2(\mathbf{r},t) = \varepsilon_0(\varepsilon_2 - 1)\mathbf{E}_2(\mathbf{r},t) + \mathbf{P}_2^{NL}(\mathbf{r},t) \qquad (2)$$

In equation (2) $\mathbf{P}_2^{NL}(\mathbf{r},t) = \varepsilon_0 \chi^{(2)} \cdot \mathbf{E}_2(\mathbf{r},t)\mathbf{E}_2(\mathbf{r},t)$ is the nonlinear polarization density due to the quadratic nonlinearity of medium 2.

For the purpose of illustration we shall consider subwavelength plasmonic spherical particles of radius $R$ randomly distributed in a dielectric host possessing a second order optical nonlinearity with a dominant term $\chi_{zzz}^{(2)}$. This choice is merely for mathematical convenience, as it is amenable to an analytical treatment. A completely analogous formulation and qualitatively similar results would hold for different particles ensembles and different host media. The specific system analyzed to produce the results reported in Figures 1, 2 and 3 is a Silver sphere of radius $R = 100nm$ embedded in a 2-methyl-4-nitroanline (MNA) host medium, which is characterized by a dominant nonlinear susceptibility term $\chi_{zzz}^{(2)} = 500\,pm/V$ [12, 13].

In the quasi-static approximation the polarization density in medium 1 can be expanded in terms of spherical harmonics $Y_{n,m}^{(e/o)}(\theta,\phi)$, defined and normalized as in [11]:

$$\mathbf{P}_1(t) = \sum_{n,m} \nabla \left\{ \frac{r^n}{R^{n-1}} \left[ P_{n,m}^{(e)}(t) Y_{n,m}^{(e)}(\theta,\phi) + P_{n,m}^{(o)}(t) Y_{n,m}^{(o)}(\theta,\phi) \right] \right\} \qquad (3)$$

Performing similar expansions for all field quantities in terms of spherical harmonics and applying the appropriate boundary conditions at the particle's interface yields the following evolution equation for the polarization density amplitude associated with any of the electromagnetic angular momentum eigenmodes of the sphere:

$$\frac{d^2 P_{n,m}^{(e,o)}(t)}{dt^2} + \gamma \frac{dP_{n,m}^{(e,o)}(t)}{dt} + \omega_n^2 P_{n,m}^{(e,o)}(t) = \omega_n^2 \frac{S_{n,m}^{(e,o)}(t)}{n} \qquad (4)$$

In equation(4), the parameter $\omega_n$ is the resonant frequency of the eigenmodes of order $n$ in the absence of nonlinear interactions and is given by:

$$\omega_n = \sqrt{\frac{n\omega_{pl}^2}{n\varepsilon_\infty + (n+1)\varepsilon_2}} \qquad (5)$$

The term $S_{n,m}^{(e,o)}(t)$ in the right-hand side of equation (4) is the projection on the spherical harmonic $Y_{n,m}^{(e,o)}(\theta,\phi)$ of the nonlinear polarization density $\mathbf{P}_2^{NL}$ evaluated over the surface of the sphere, i.e.:

$$S_{n,m}^{(e,o)}(t) = \oint_{r=R} Y_{n,m}^{(e,o)}(\theta,\phi) \mathbf{P}_2^{(NL)}(R,\theta,\phi,t) \cdot \hat{\mathbf{r}} \sin(\theta) d\theta d\phi \qquad (6)$$

As a consequence of the term (6) various eigenmodes are nonlinearly coupled to each other and to the pump field. The symmetry group of medium 2, along with the spatial profile of the pump, determine which specific three-wave mixing products contribute to the dynamics of a given eigenmode.

Proceeding by way of example, we consider the dynamics of the azimuthally uniform ($m = 0$) resonant mode of order $n$ in the presence of a spatially uniform pump. In this instance the evolution equation (4) assumes the following form:

$$\frac{d^2 P_{n,0}^{(e)}(t)}{dt^2} + \gamma \frac{dP_{n,0}^{(e)}(t)}{dt} + \omega_n^2 P_{n,0}^{(e)}(t) =$$
$$= \alpha_1 E_P(t) P_{n,0}^{(e)}(t) + \alpha_2 \left[ P_{n,0}^{(e)}(t) \right]^2 \qquad (7)$$

The first term in the right hand side of equation (7) represents the three-wave mixing process responsible for the parametric interaction of the resonant mode with the pump. The second term in the right hand side of equation (7) accounts the sum and difference frequency generation processes due to the plasmonic mode itself interacting with the background medium. The expressions of the nonlinear interaction coefficients $\alpha_1$ and $\alpha_2$ are given by:

$$\alpha_1 = -\frac{4\sqrt{\pi}\chi_{zzz}}{n\sqrt{3}} \frac{\omega_n^4}{\omega_{pl}^2} G_{n,0}^{(e,e)} \;;\; \alpha_2 = \frac{\chi_{zzz}}{n\varepsilon_0} \frac{\omega_n^6}{\omega_{pl}^4} F_{n,0}^{(e,e,e)}$$

$$F_{n,0}^{(e,e,e)} = \int_0^{2\pi} \int_0^{\pi} \left\{ \cos\theta \frac{\partial}{\partial z} \left[ \frac{R^{n+2}}{r^{n+1}} Y_{n,0}^{(e)}(\theta,\phi) \right]_R \times \right.$$
$$\left. \times \frac{\partial}{\partial z} \left[ \frac{R^{n+2}}{r^{n+1}} Y_{n,0}^{(e)}(\theta,\phi) \right]_R Y_{n,0}^{(e)}(\theta,\phi) \right\} \sin\theta\, d\theta d\phi \qquad (8)$$

$$G_{n,0}^{(e,e)} = \int_0^{2\pi} \int_0^{\pi} \left\{ \cos\theta \frac{\partial}{\partial z} \left[ r Y_{1,0}^{(e)}(\theta,\phi) \right]_R \times \right.$$
$$\left. \times \frac{\partial}{\partial z} \left[ \frac{R^{n+2}}{r^{n+1}} Y_{n,0}^{(e)}(\theta,\phi) \right]_R Y_{n,0}^{(e)}(\theta,\phi) \right\} \sin\theta d\theta d\phi$$

The PPR threshold is minimized [11] if the system is driven at the second harmonic frequency of the mode of interest. We start therefore by considering a spatially uniform monochromatic pump field of the form $E_P(t) = A_p \sin(2\omega_n t)$. Under these conditions the linear interaction of the pump field with the subwavelength particle is essentially of electric dipolar origin. Importantly the pump field at $2\omega_n$ in the case at hand is strongly detuned from any resonant mode, including the dipolar one occurring at $\omega_1$, so a relatively small pump absorption is expected. The linear absorption cross-section of a subwavelength of volume $V$ at the pump frequency $2\omega_n$, in the absence of any nonlinear process (i.e. for $\alpha_1, \alpha_2 = 0$) can be written as follows:

$$\sigma_{2\omega_n} = \frac{36V\varepsilon_0 \eta \left(\omega_n^2 - \omega_1^2\right)^2 \omega_{pl}^2 \gamma}{(n-1)^2 \left(4\omega_n^2 - \omega_1^2\right)^2 \omega_n^2} \qquad (9)$$

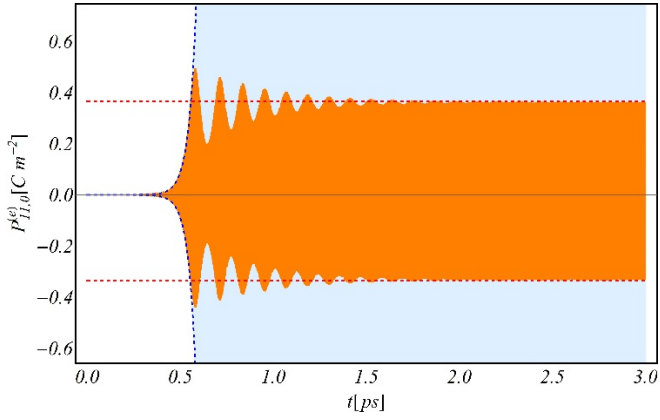

Figure 1. Polarization density amplitude term $P_{11,0}^{(e)}$ of a Silver sphere immersed in a MNA background medium. In order to better highlight the relaxation oscillations occurring in the system we show a case in which the PPR threshold is exceeded by a large margin ($A_p = 10 A_{PPR}$). The dashed lines show the oscillation limits predicted by the asymptotic formula (12).

It is straightforward to verify that, given the large detuning of the pump frequency $2\omega_n$ from the dipolar resonance $\omega_1$, equation (9) yields a very small value for the absorption cross-section – much smaller in fact than the geometric cross-section of the particle. We turn then to the nonlinear interactions to show how the absorption cross-section dramatically increases when the threshold condition for PPR is met.

In solving equation (7) we notice at the outset that so long as the condition $|P_{n,m}^{(e,o)}(t)| \ll \alpha_1 A_p / \alpha_2$ holds, which is certainly the case at least at the initial stages of the parametric interaction, a solution can be easily obtained in terms of Mathieu functions [14, 15]. More physically transparent though is the following slowly-varying-envelope approximate solution:

$$P_{n,m}^{(e,o)}(t) = p(t) \cos[\omega_n t - \theta(t)] e^{-\frac{\gamma}{2} t}$$

$$p(t) = p_0 \sqrt{\cosh\left(\frac{\alpha_1 A_p}{2\omega_n} t\right)} \; ; \; \theta(t) = \operatorname{arccot}\left[\exp\left(-\frac{\alpha_1 A_p}{2\omega_n} t\right)\right]$$
(10)

where $p_0$ is the initial modal amplitude.

From equation (10) for $p(t)$ it is apparent that the system enters the PPR regime provided that the pump electric field amplitude $A_p$ exceeds the threshold value:

$$A_{PPR} = 2\gamma \omega_n / \alpha_1 \qquad (11)$$

Such threshold condition separates distinct dynamics, whereby $P_{n,m}^{(e,o)}(t)$ decreases exponentially for $A_p < A_{PPR}$ and increases exponentially for $A_p > A_{PPR}$. Such contrasting modal dynamics are reflected in the distinct absorption regimes that the pump experiences. The power parametrically transferred from the pump to the resonant mode is given by:

$$W_{abs}(t) = \frac{n R^3 \alpha_1 A_p p_0^2}{32 \varepsilon_0 \omega_{pl}^2} \left[2\omega_n + \gamma \sinh\left(\frac{\alpha_1 A_p t}{2\omega_n}\right)\right] e^{-\gamma t}$$

$$\sim \frac{n R^3 \alpha_1 A_p p_0^2 \gamma}{64 \varepsilon_0 \omega_{pl}^2} \exp\left[(A_p - A_{PPR}) \frac{\alpha_1 t}{2\omega_n}\right]$$
(12)

Equation (12) highlights the fundamental trait of PPA which is stark contrast with linear absorption: in PPA absorption is vanishingly small for incident fields below the PPR threshold, and increases exponentially under high intensity excitation.

Clearly a saturation of the exponential behavior is expected, because, if nothing else, the absorbed power (12) cannot exceed the finite power carried by the pump. In fact a different mechanism limits (12) before pump depletion occurs. Such mechanism is the resonance detuning due to the last quadratic term in equation (7) that we have neglected in our analysis so far. As $|P_{n,m}^{(e,o)}(t)| \sim \alpha_1 A_p / \alpha_2$ equation (7) can only be integrated numerically. Nevertheless the following asymptotic expressions as $t \to \infty$ for $P_{n,m}^{(e,o)}(t)$ hold for $A_p > A_{PPR}$:

$$P_{n,m}^{(e,o)}(t) \sim -\frac{\alpha_2 Q_1^2}{2\omega_n^2} + Q_1 \cos(\omega_n t + \theta_1) + \frac{\alpha_2 Q_1^2}{6\omega_n^2} \cos(2\omega_n t + 2\theta_1)$$

$$\theta_1 = \frac{1}{2} \arccos\left(-\frac{A_{PPR}}{A_p}\right) \; ; \; Q_1 = \frac{\omega_n}{\alpha_2} \sqrt{\frac{6\gamma \omega_n}{5}} \sqrt{\left(\frac{A_p}{A_{PPR}}\right)^2 - 1}$$
(13)

Within the range of validity of equations (13) the exponentially growing oscillations of the polarization density amplitude level-off as $t \to \infty$ after a sequence of relaxation oscillations. In figure 1 the numerical solution of equation (7) (indicated in orange) is compared with the predictions of the analytical model (10), shown in blue, and representing the PPR dynamics in the absence of saturation effects (i.e. $\alpha_2 = 0$). As evident from this comparison the PPR model (10) faithfully reproduces the dynamics of the actual system until the polarization density amplitude of the mode attains values close to asymptotic estimate (13), shown by the red dashed lines. At that point the exponential trend transitions into a train of amplitude oscillations that eventually relax towards a constant steady state.

Based on (13) the average power $\bar{W}_p$ transferred from the pump to the plasmonic mode approaches asymptotically the value:

$$\bar{W}_{abs}(t \to \infty) = \frac{3}{20} n R^3 \frac{\gamma}{\varepsilon_0 \omega_{pl}^2} \frac{\omega_n^3}{\alpha_2^2} \sqrt{\left(\frac{A_p}{A_{PPR}}\right)^2 - 1} \qquad (14)$$

Using the steady state asymptotic estimate (14) of the absorbed power it is possible to obtain the PPR contribution to the particle absorption cross-section (in addition to the linear portion given by equation(9)):

$$\sigma_{NL} = \frac{3nR^3}{40\varepsilon_0 \sqrt{\varepsilon_2}} \frac{\omega_n^3}{\omega_{pl}^2} \frac{\alpha_1^2}{\alpha_2^2} \sqrt{\frac{I_{PPR}}{I_p}\left(1 - \frac{I_{PPR}}{I_p}\right)} \; , \; I_p > I_{PPR} \qquad (15)$$

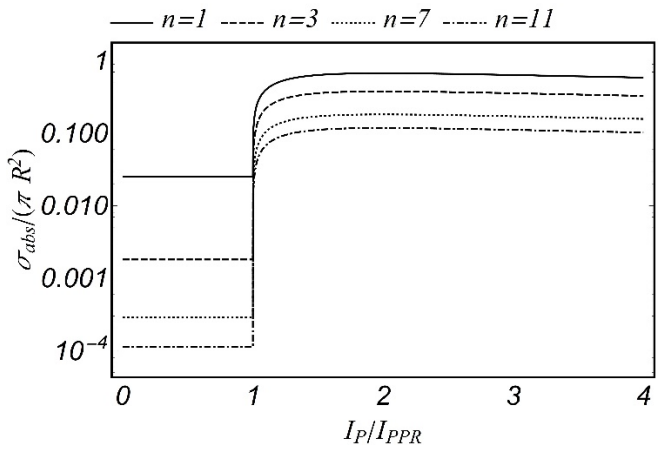

Figure 2. Absorption cross-section of the plasmonic particle normalized to the geometrical cross-section, as a function of the incident intensity.

In equation (15) $I_p = A_p^2/(2\eta)$ is the incident intensity and $I_{PPR} = A_{PPR}^2/(2\eta)$ is the PPR intensity threshold, where $\eta$ is the characteristic impedance of the background medium. If particles similar to the one described so far are dispersed with density $N$ in the background medium, the nonlinear pump attenuation coefficient of the composite follows from equation (15) as $\alpha_{NL} = N\sigma_{NL}$.

Figure 2 shows how the normalized absorption cross-section of a silver sphere or radius $R = 100 nm$ in a MNA host is affected by various modes undergoing PPR. The absorption cross-section is plotted against the normalized pump intensity and, for each of the PPR modes considered in figure 2, the incident pump field is at twice the value of the corresponding resonant frequency given by equation (5). For the case at hand all the possible PPR resonant wavelengths $\lambda_n$ fall in the range $\lambda_\infty < \lambda_n \le \lambda_1$, where $\lambda_\infty = 563 nm$ and $\lambda_1 = 448 nm$. As evident from figure 2, as soon as a pump field of frequency $2\omega_n$ exceeds the intensity threshold $I_{PPR}$ the particle's absorption cross-section increases dramatically due to the contribution of the mode of eigenfrequency $\omega_n$ undergoing PPR. This phenomenon is form of reverse saturable absorption and could have interesting applications in optical limiting devices, especially given the design versatility of metallic nanoparticles for targeting different spectral regions.

For the purpose of illustration in figure 3 we apply the analysis and the models developed thus far to the practically relevant case of a pulsed pump. The pulse considered here is a 40ps pulse of average power 0.5W, focused to an area of $25\mu m^2$, on parametric resonance with the n=11 m=0 mode ($\lambda_{11} = 460 nm$) of a silver particle of radius 100nm embedded in a MNA background medium. The red curve in figure 3 shows the total instantaneous power of the pump field. The blue curve shows the instantaneous absorbed power mediated by the PPR process. The vertical dashed lines demarcate the time interval in which the pump field exceeds the PPR threshold. A similar behavior is observed both in figure 1 and figure 3 at the onset of PPR, where the modal polarization (and the corresponding absorption) builds up exponentially to slightly surpass the steady-state value, and then relaxes to such value through a series of oscillations (only one is discernible in figure 3).

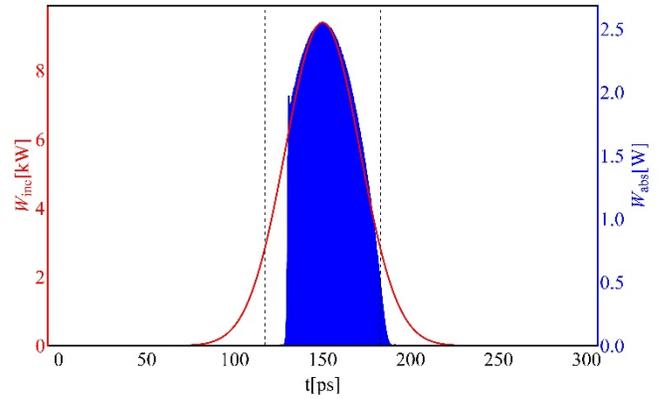

Figure 3. The red curve shows the instantaneous pump power $W_{inc}$. The blue curve show the instantaneous pump power $W_{abs}$ absorbed by the particle via PPR of the n=11, m=0 mode. The dashed vertical lines indicate the times at which the pump intensity is equal to the PPR threshold.

Figure 3 confirms the dramatic increase in absorption that PPAs exhibit under high intensity illumination.

In conclusion we have introduced the theory of Plasmonic Parametric Absorbers (PPA). In particular we have shown that PPAs exhibit a reverse saturable absorption behavior whereby an incident field that is parametrically resonant with one or more of the modes of a plasmonic particle experiences a strongly enhanced absorption whenever its intensity exceeds the relevant PPR threshold. Such effect makes PPAs very promising candidates for optical limiting applications, in addition of being of fundamental interest in the emerging field of nonlinear plasmonics.

**Aknowledgements**
A.S. acknowledges the support of the Air Force Office of Scientific Research (AFOSR) through the grant FA9550-16-1-0152.

**References**
1. N. Landy, S. Sajuyigbe, J. Mock, D. Smith, and W. Padilla, Phys. Rev. Lett. **100**, 207402 (2008).
2. C. M. Watts, X. Liu, and W. J. Padilla, Advanced materials **24** (2012).
3. H. Tao, N. I. Landy, C. M. Bingham, X. Zhang, R. D. Averitt, and W. J. Padilla, Opt. Express **16**, 7181-7188 (2008).
4. J. Hao, J. Wang, X. Liu, W. J. Padilla, L. Zhou, and M. Qiu, Appl. Phys. Lett. **96**, 251104 (2010).
5. X. Shen, T. J. Cui, J. Zhao, H. F. Ma, W. X. Jiang, and H. Li, Opt. Express **19**, 9401-9407 (2011).
6. F. Ding, Y. Cui, X. Ge, Y. Jin, and S. He, Appl. Phys. Lett. **100**, 103506 (2012).
7. J. Grant, Y. Ma, S. Saha, A. Khalid, and D. R. Cumming, Opt. Lett. **36**, 3476-3478 (2011).
8. Y. Avitzour, Y. A. Urzhumov, and G. Shvets, Phys. Rev. B **79**, 045131 (2009).
9. X. Liu, T. Starr, A. F. Starr, and W. J. Padilla, Phys. Rev. Lett. **104**, 207403 (2010).
10. J. Ng, H. Chen, and C. T. Chan, Opt. Lett. **34**, 644-646 (2009).
11. A. Salandrino, Phys. Rev. B **97**, 081401 (2018).
12. D. S. Chemla, *Nonlinear optical properties of organic molecules and crystals* (Elsevier, 2012).
13. M. Mayy, G. Zhu, A. Webb, H. Ferguson, T. Norris, V. Podolskiy, and M. Noginov, Opt. Express **22**, 7773-7782 (2014).
14. E. Mathieu, Journal de mathématiques pures et appliquées, 137-203 (1868).
15. G. B. Arfken, *Mathematical methods for physicists* (Academic press, 2013).

**Extended References**

1. N. Landy, S. Sajuyigbe, J. Mock, D. Smith, W. Padilla, Perfect metamaterial absorber, Phys. Rev. Lett., 100 (2008) 207402.

2. C.M. Watts, X. Liu, W.J. Padilla, Metamaterial electromagnetic wave absorbers, Advanced materials, 24 (2012).

3. H. Tao, N.I. Landy, C.M. Bingham, X. Zhang, R.D. Averitt, W.J. Padilla, A metamaterial absorber for the terahertz regime: design, fabrication and characterization, Opt. Express, 16 (2008) 7181-7188.

4. J. Hao, J. Wang, X. Liu, W.J. Padilla, L. Zhou, M. Qiu, High performance optical absorber based on a plasmonic metamaterial, Appl. Phys. Lett., 96 (2010) 251104.

5. X. Shen, T.J. Cui, J. Zhao, H.F. Ma, W.X. Jiang, H. Li, Polarization-independent wide-angle triple-band metamaterial absorber, Opt. Express, 19 (2011) 9401-9407.

6. F. Ding, Y. Cui, X. Ge, Y. Jin, S. He, Ultra-broadband microwave metamaterial absorber, Appl. Phys. Lett., 100 (2012) 103506.

7. J. Grant, Y. Ma, S. Saha, A. Khalid, D.R. Cumming, Polarization insensitive, broadband terahertz metamaterial absorber, Opt. Lett., 36 (2011) 3476-3478.

8. Y. Avitzour, Y.A. Urzhumov, G. Shvets, Wide-angle infrared absorber based on a negative-index plasmonic metamaterial, Phys. Rev. B, 79 (2009) 045131.

9. X. Liu, T. Starr, A.F. Starr, W.J. Padilla, Infrared spatial and frequency selective metamaterial with near-unity absorbance, Phys. Rev. Lett., 104 (2010) 207403.

10. J. Ng, H. Chen, C.T. Chan, Metamaterial frequency-selective superabsorber, Opt. Lett., 34 (2009) 644-646.

11. A. Salandrino, Plasmonic parametric resonance, Phys. Rev. B, 97 (2018) 081401.

12. D.S. Chemla, Nonlinear optical properties of organic molecules and crystals, Elsevier, 2012.

13. M. Mayy, G. Zhu, A. Webb, H. Ferguson, T. Norris, V. Podolskiy, M. Noginov, Toward parametric amplification in plasmonic systems: Second harmonic generation enhanced by surface plasmon polaritons, Opt. Express, 22 (2014) 7773-7782.

14. E. Mathieu, Mémoire sur le mouvement vibratoire d'une membrane de forme elliptique, Journal de mathématiques pures et appliquées, (1868) 137-203.

15. G.B. Arfken, Mathematical methods for physicists, Academic press, 2013.